\documentclass[sigconf,nonacm]{acmart}
\AtBeginDocument{%
  }

\usepackage[most]{tcolorbox}

\newtcolorbox{chatlog}[1]{
  colback=gray!5,
  colframe=black,
  boxrule=0.5pt,
  arc=2pt,
  left=6pt,
  right=6pt,
  top=5pt,
  bottom=5pt,
  title=\textbf{#1},
  fonttitle=\small,
  before skip=8pt,
  after skip=8pt
}

\usepackage{tabularx}
\usepackage{multirow}
\newcolumntype{Y}{>{\hsize=.6\hsize}X}    
\newcolumntype{W}{>{\hsize=.7\hsize}X}    
\newcolumntype{V}{>{\hsize=.85\hsize}X}   
\newcolumntype{Z}{>{\hsize=2.25\hsize}X}  
\usepackage{verbatim}
\usepackage{booktabs}

\begin{document}


\title[Verify, Offload, Extend \& Recommend: Selective Complementarity in AI Support]{Verify, Offload, Extend \& Recommend: Selective Complementarity in AI Support for Physical Activity Planning with Longitudinal Patient Data}


\author{Pavithren V. S. Pakianathan}
\affiliation{%
  \institution{Ludwig Boltzmann Institute for Digital Health and Prevention}
  \city{Salzburg}
  \country{Austria}
}
\affiliation{%
  \institution{LMU Munich}
  \city{Munich}
  \country{Germany}
}

\author{Rania Islambouli}
\affiliation{%
  \institution{Ludwig Boltzmann Institute for Digital Health and Prevention}
  \city{Salzburg}
  \country{Austria}
}

\author{Diogo Branco}
\affiliation{%
  \institution{LASIGE, Faculdade de Ciências, Universidade de Lisboa}
  \city{Lisbon}
  \country{Portugal}
}

\author{Gil Rosa}
\affiliation{%
  \institution{Faculty of Human Kinetics, Universidade de Lisboa}
  \city{Lisbon}
  \country{Portugal}
}

\author{Rita Pinto}
\affiliation{%
  \institution{CCUL@RISE, Faculdade de Medicina, Universidade de Lisboa}
  \city{Lisbon}
  \country{Portugal}
}

\author{Albrecht Schmidt}
\affiliation{%
  \institution{LMU Munich}
  \city{Munich}
  \country{Germany}
}

\author{Tiago Guerreiro}
\affiliation{%
  \institution{LASIGE, Faculdade de Ciências, Universidade de Lisboa}
  \city{Lisbon}
  \country{Portugal}
}

\author{Jan David Smeddinck}
\affiliation{%
  \institution{Ludwig Boltzmann Institute for Digital Health and Prevention}
  \city{Salzburg}
  \country{Austria}
}
\renewcommand{\shortauthors}{V S Pakianathan et al.}

\begin{abstract}

Self-tracking technologies generate longitudinal patient-generated health data that can provide insight into everyday behavior, but interpreting these data alongside clinical information remains a demanding sensemaking task for healthcare professionals (HCPS).We investigate how generative AI can support this process in physical activity planning for cardiovascular disease. In a counterbalanced within-subjects study, 26 HCPs developed plans for four real patient cases using longitudinal wearable data, with and without AI-generated summaries and conversational support, followed by an exploratory evaluation of an AI exercise-plan generator. AI contributions varied across HCPs and tasks. Visualization literacy moderated effects on expert-rated plan quality and perceived workload, suggesting different forms of benefit for different users. Across interviews and 152 chatbot queries, we identified four roles for AI support: verifying interpretations, offloading information-processing work, extending professional knowledge, and generating editable plan starting points. We characterize these findings as selective complementarity, with professional judgment remaining central to final decisions.

\end{abstract}




\maketitle

\section{Introduction}

Healthcare professionals (HCPs) increasingly make decisions using information collected beyond the clinical encounter~\cite{ye_impact_2021}. This is particularly relevant in cardiovascular disease (CVD) -- the largest cause of death worldwide ~\cite{who_cardiovascular_nodate}-- prevention and rehabilitation, where regular physical activity is a key component of cardiovascular risk reduction and long-term disease management~\cite{pinckard_effects_2019,back_2026_2026,bull_world_2020}. Understanding how patients are active in everyday life is therefore important for informing physical activity planning and ongoing risk management~\cite{mcgowan_exploring_2024}. Patient-generated health data (PGHD), including longitudinal measures from wearable technologies, can provide insight into everyday patient behavior and support more personalized care~\cite{branco_co-designing_2024,taxter_co-design_2022}. For physical activity planning, these longitudinal data can help HCPs understand habitual activity patterns that need to be considered alongside cardiovascular conditions, medications, physical limitations, exercise preferences, and treatment goals~\cite{pakianathan_towards_2026,pakianathan_dataenabledDialogue_2026,taylor_practical_2023}.

Clinical dashboards can support physical activity planning by bringing longitudinal PGHD and clinical information together and making patterns in patient data more accessible~\cite{branco_co-designing_2024,sutton2020overview,taxter_co-design_2022}. However, HCPs still need to make sense of information distributed across different data sources and representations. For example, when developing a physical activity plan, an HCP may need to examine activity patterns across several weeks, determine whether activity levels are consistent or changing, relate these patterns to exercise preferences and physical limitations, and consider whether medications or cardiovascular conditions require additional precautions~\cite{taylor_practical_2023,back_2026_2026}. This involves not only inspecting individual data points or visualizations, but synthesizing them into an overall understanding of the patient and translating that understanding into an appropriate plan. As the amount and diversity of patient information increase, this synthesis can create additional information-processing demands~\cite{nijor_patient_2022}.

Generative AI offers new ways of dynamically supporting parts of this sensemaking process while retaining access to the underlying patient data and visualizations. Large Language Models (LLMs) can summarize information distributed across multiple sources and support natural-language exploration of patient data~\cite{bednarczyk_scientific_2025}. For example, an AI-generated summary could provide an overview of activity trends that an HCP might otherwise need to manually derive from several visualizations, while a conversational interface could also allow the HCP to ask targeted questions about exercise frequency, changes over time, or the implications of clinical information for physical activity planning. In this way, generative AI can be considered a complementary resource for navigating and synthesizing patient information rather than a replacement for conventional data representations.

Introducing AI into clinical workflows, however, also changes the way information is encountered and processed. AI-generated summaries, responses, and recommendations can influence what information professionals attend to and how they combine system output with their own expertise. Previous work on AI-supported clinical decision support has raised questions around workflow integration, appropriate reliance, explainability, and professional agency~\cite{abreu_ai_2026,rajashekar_human-algorithmic_2024,yang_unremarkable_2019,zhu_augmenting_2026}. Much of this research has focused on AI systems that provide predictions, diagnoses, risk estimates, or recommendations ~\cite{nojomi2025ai}. Less is known about how generative AI is integrated into open-ended planning tasks in which HCPs must interpret longitudinal behavioral data, combine it with clinical and contextual information, and construct a personalized plan. In such settings, AI becomes one of several resources available during sensemaking, alongside visualizations, patient information, and professional knowledge. Understanding its role therefore requires examining both how AI support relates to planning outcomes and how HCPs incorporate it into the planning process.

To investigate this, we conducted a counterbalanced within-subjects study with 26 exercise physiologists who developed physical activity plans for four cardiovascular cases using patient-donated longitudinal patient-generated health datasets. Participants worked on two cases using a decision-support interface presenting clinical information and longitudinal data visualizations, and two cases using the same interface augmented with AI-generated summaries and conversational support. Condition order was counterbalanced such that half the participants received the two AI-assisted cases first, and half received them last. Additionally, the sequence of the four individual patient cases was permuted across participants using four distinct presentation orders to control for case-specific sequence effects. In both conditions, HCPs interpreted the available information and constructed the final physical activity plans themselves. We combined measures of workload, usability, plan confidence, and independent expert ratings of plan quality with interviews and interaction-log analysis to examine both planning outcomes and how AI was incorporated into the sensemaking and planning process. We address the following research questions:

\textbf{RQ1:} How does AI support shape decision-making in physical activity planning and how does its impact vary across HCPs.

\textbf{RQ2:} How and when do HCPs use AI support during clinical decision-making for physical activity planning and data sensemaking?

Our findings reveal a heterogeneous pattern in the efficacy of AI support in physical activity planning. The effects of AI varied with visualization literacy, while HCPs appropriated AI differently across the planning process, using it to \textbf{verify} interpretations, \textbf{offload} information-processing work, and \textbf{extend} immediate knowledge. Together, these findings point to \emph{selective complementarity}: AI support becomes valuable when particular AI capabilities align with the needs and capabilities of the professional at a given point in the task. Rather than implying uniform benefits, this perspective highlights how different forms of AI support can complement  clinical decision-making and professional judgment in different ways. Building on these findings, we characterize selective complementarity as a form of human–AI collaboration in which AI contributes by complementing professional capabilities rather than providing the same benefit across users and contexts. We discuss implications for designing AI-supported decision support systems that preserve clinical decision-making and professional judgment.

\section{Related Work}
\subsection{Integrating Clinical and Patient-Generated Health Data into Clinical Workflows} 
Healthcare decision-making is faced with an increasing opportunity to incorporate data from traditional clinical assessments and electronic health records as well as patient-generated health data collected through self-tracking technologies~\cite{ye_impact_2021}. To monitor and prevent diseases, HCPs may use visualization dashboards which help provide insights for decision-making\cite{pakianathan_towards_2026}. However, data-integration should be designed well to align with clinical workflows. Branco et al.~\cite{branco_co-designing_2024} highlight that not having enough data or having unnecessary data could result in non-adoption. On the other hand, when there is too much data, it could result in additional burden for clinicians. Ye et al.~\cite{ye_impact_2021} suggest that technostress, workflow-related issues and time pressure issues arise when integrating PGHD into EHR resulting in increased risk of burnout and that AI could help automate interpretation of large amounts of data into digestible sizes and assists with decision support. Similarly, the American Heart Association suggested that automation -- e.g. using AI to automate interpretation of clinical and digital data to assist clinicians and also fully automate low-value tasks -- could help to reduce the burden on clinicians and minimize risk of fatigue and burnout~\cite{golbus_digital_2023}. Self-tracked data by cardiac patients such as physical activity, blood pressure, and sleep and steps can be highly valuable for assessing risk, physical activity adherence, personalizing treatment~\cite{nick_e_j_west_personalized_2022,tadas_using_2023,back_2026_2026}.
However, limited research exists on how patient generated health data can be integrated into cardiac prevention workflows. While some research protocols in similar directions have been published \cite{ventura_clinical_2022}, there are no established decision-support tools in the context of CVD physical activity planning workflows that integrate patient generated health data to date.

\subsection{AI-Enabled Clinical Decision Support Systems} 

Clinical decision support systems (CDSS) support healthcare professionals in integrating patient information, clinical evidence, and guidelines during decision-making. Traditionally, CDSS have included knowledge-based systems using predefined rules as well as data-driven approaches; recent advances in AI and LLMs have expanded their ability to synthesize heterogeneous information, answer clinical questions, and generate recommendations~\cite{sutton2020overview,elhaddad_ai-driven_2024}. However, successful integration depends not only on model capabilities but also on how these systems fit clinical workflows, support professional reasoning, and appropriate reliance~\cite{rajashekar_human-algorithmic_2024,yang_unremarkable_2019}. To address this researchers have suggested that the human-element and HCI considerations are of utmost importance. HCI was cited to be the top challenge for CDSS integration~\cite{sittig_grand_2008} and researchers, argue that CDSS integration is a socio-technical challenge~\cite{rajashekar_human-algorithmic_2024,jacobs2021designing}. Both Yang et al.~\cite{yang_unremarkable_2019} and Elhaddad et al.~\cite{elhaddad_ai-driven_2024} emphasize that HCI considerations, user-centered design, workflow integration are key for successful CDSS integration. Similarly researchers emphasize a need for further research on how HCPs interact with AI throughout the entire clinical decision-making process~\cite{abreu_ai_2026}. At a healthcare system level, the Clinical Decision Support Innovation Collaborative (CDSiC) Implementation, Adoption, and Scaling Workgroup in the USA, highlights the need for clinical involvement developing in AI CDSS ~\cite{kawamoto2024implementation} for better contextual fit.

Recent CHI work has explored AI(LLM) integration in CDSS. LLMs' abilities to answer clinical questions and summarizing information has increased its feasibility of use for decision making in general medicine and radiology and pediatrics~\cite{rajashekar_human-algorithmic_2024}. Zhu et al.~\cite{zhu_augmenting_2026} designed AIcare, an AI copilot for collaborative clinical decision-making powered by LLMs. They found it to reduce workload for HCPs overall and that for junior HCPs it helped with cognitive scaffolding and for senior HCPs, it served as a tool for adversarial verification. In another work, Rajashekar et al.~\cite{rajashekar_human-algorithmic_2024}, designed an AI-CDSS for the management of upper gastrointestinal bleeding and found that LLM augmentation improved ease of use and that HCI varied based on level of clinical expertise. Both studies found AI-CDSS to serve as copilots or team partners used to augment the HCP judgment.

Building on the potential of LLMs to summarize heterogeneous data~\cite{li_vital_2025} and support natural-language exploration and question-answering~\cite{demner-fushman_what_2009}, we investigate how such capabilities can support clinical decision-making when HCPs integrate longitudinal PGHD with clinical information during physical-activity planning in cardiac rehabilitation. Prior work has shown that LLM-based copilots can support collaborative clinical decision-making in domains such as nephrology and obstetrics~\cite{zhu_augmenting_2026}. Our study extends this line of inquiry into clinical-decision making from longitudinal behavioral and clinical information.

\subsection{Human-AI Collaboration in AI-CDSS}
Researchers argue that AI-CDSS should be designed well for workflow and contextual integration and sustained use by clinicians~\cite{jacobs2021designing,panigutti_co-design_2023,rajashekar_human-algorithmic_2024}. This could allow for optimal Human-AI collaboration and help unlock complementary team performance ~\cite{bansal_does_2021} -- leveraging AI and human capabilities to achieve better performance compared to either one performing the task alone ~\cite{hemmer_complementarity_2025}. Rajashekhar et al. emphasize that "AI-CDSS should be designed as interactive systems that physicians can use to support their cognitive processes, as part of a human-AI collaboration paradigm."~\cite{rajashekar_human-algorithmic_2024}. Aligned with this, Yang et al. ~\cite{yang_unremarkable_2019}, by following the concepts of unremarkable computing by Tolmie et al.~\cite{tolmie_unremarkable_2002} strive to make AI 'invisible in use' while designing a CDSS -- passively situated in existing routines and being noticed only when it could add value to a decision-making routine. In a hospital or clinical environment, a deluge of information from electronic health records and other data sources such as PGHD could overwhelm HCPs~\cite{ye_impact_2021}. In this context, from an HCI perspective, by applying the theory of distributed cognition~\cite{hollan_distributed_2000}, AI can help clinicians by offloading cognition for decision-making tasks particularly involving sensemaking of data.

However, identifying the right strategy to implement Human-AI collaboration in AI CDSS is challenging and highly contextual~\cite{zajac_it_2024}. Studies have found heterogenous results with AI integration for decision-support~\cite{gong_cognitive_2026,abreu_ai_2026,rajashekar_human-algorithmic_2024,heudel_artificial_2026}. Gong et al. ~\cite{gong_cognitive_2026} in their systematic review found diagnostic and alerting AI tools showed mixed effects on cognitive workload and were highly dependent on implementation characteristics, alert frequency, case complexity, and workflow integration. Similarly, in a study focusing on emergency room clinical cases, Abreu et al.~\cite{abreu_ai_2026} found that for easier cases, AI could help HCPs with selecting appropriate tests and reduce decision-making time, but for complex cases, although diagnostic accuracy was improved, it also led to higher medical resource use where HCPs could overprescribe tests, including unnecessary ones. Conversely, in a study trial deploying an AI-CDSS for the management of upper gastrointestinal bleeding~\cite{rajashekar_human-algorithmic_2024} AI improved usability and that level of clinical expertise influenced interaction patterns. Overall, there is a need for understanding the contextual fit for integrating AI to support Human-AI collaboration for clinical decision-making.

Physical activity planning in cardiac rehabilitation is a unique context which involves looking at the patients' physical activity, their risk factors, their medications, activity preferences, physical limitations which makes decision-making complex. Our goal was to investigate how future decision support tools integrating patient generated health data and electronic health record data augmented with AI support -- including generated summaries and a chatbot -- into a clinical dashboard reshapes healthcare professionals’ sensemaking and physical-activity planning. This was driven by our motivation to model better Human-AI collaboration. We examine whether AI support affects workload, confidence, usability, and externally assessed plan quality, also how HCPs appropriate AI during decision-making and how its value varies with user characteristics and case demands. 
\section{Method}

To investigate how AI support shapes decision-making in physical-activity planning and understand how HCPs appropriate AI during decision-making, we conducted a within-subjects study comparing an AI-assisted condition against a No-AI baseline CDSS (Figure \ref{fig:wireframe}) with four real and anonymized patient cases integrating patient-donated PGHD. The study was conducted in the [BLINDED] language.

\subsection{Patient Cases and Data Collection}
To ensure ecological validity, we utilized four clinical patient profiles using real-world self-tracking data donated by four patients (P1–P4) with cardiovascular conditions for the purposes of this study. The donor patients had continuously tracked their vitals using consumer wearables (Garmin, $n=2$; Apple Watch, $n=2$) for over one year. We retained only physical activity measures used by the decision-support interface: daily steps, moderate and vigorous physical activity (MPA/VPA), and recorded activity sessions (activity type and duration). These measures were informed by prior literature \cite{pakianathan_towards_2026} and consultation with the co-authors specializing in sports science and cardiac rehabilitation. Data from Apple Watch and Garmin were harmonized into a common daily- and session-level schema. Garmin MPA/VPA minutes were obtained from Garmin Connect, whereas Apple Watch MPA/VPA minutes were derived from heart-rate data using age-based intensity thresholds; step counts were aggregated by day, and recorded physical activities were standardized to physical activity type and duration in minutes.

The clinical characteristics, medication regiment, and physical activity profiles of these cases are summarized in Table~\ref{tab:patient_demographics}. Patients ranged in age from 51 to 62 years ($M = 57.8$, $SD = 4.7$; 3 male, 1 female), all had underlying Coronary Artery Disease (CAD), and all were categorized as physically active based on baseline RAPA scores (scores 6–7). Although we did not control the case complexity in the experiment we had some differences in within the patients. While P1, P3, and P4 presented moderate clinical complexity (1–5 concurrent cardiovascular medications), Patient case 2 presented considerably higher clinical complexity due to comorbid Type 1 Diabetes Mellitus and a complex daily regimen consisting of eight medications.

\begin{table*}[t]
\centering
\caption{Case Details of Donor Patient including Clinical and Behavioral details and Medication Regimen}
\label{tab:patient_demographics}
\small
\begin{tabularx}{\textwidth}{@{}c@{\hspace{4pt}}c@{\hspace{4pt}}c@{\hspace{4pt}}c@{\hspace{6pt}}W Y V c Z Y c@{}}
\toprule
\textbf{ID} & \textbf{Age/Sex} & \textbf{BMI} & \textbf{BP} & \textbf{Clinical Conditions} & \textbf{Wearable} & \textbf{Preferred Activities} & \textbf{RAPA} & \textbf{Observations \& Limitations} & \textbf{Dosing Windows} & \textbf{Meds ($n$)} \\ \midrule
1 & 51 / M & 28.7 & 121/59 & CAD & Apple Ultra & Walking, Elliptical & 7 & Back problems (unable to run); heat sensitivity & Morning, Night & 3 \\ \addlinespace
2 & 59 / F & 29.4 & 128/79 & CAD, T1D & Garmin Vénus 3 & Walking, Gym Training & 6 & T1D since age 16; sedentary at work; MD-recommended wearable & Morning, Lunch, Night & 8 \\ \addlinespace
3 & 62 / M & 25.7 & 98/64  & CAD, HF & Apple Series 9 & Walking, Gym Training & 6 & High stress / busy schedule; wife-supported HR tracking & Morning, Lunch, Night & 5 \\ \addlinespace
4 & 59 / M & 25.2 & 126/90 & CAD & Garmin Fénix 7 Pro & Cycling, Mountain Biking & 6 & Irregular schedule; compensates with $>70$km weekend bike rides & Morning, Lunch, Night & 3 \\ \bottomrule
\end{tabularx}
\vskip 4pt
\caption*{{\footnotesize \textit{Note:} BP = Blood Pressure (Systolic/Diastolic in mmHg); CAD = Coronary Artery Disease; HF = Heart Failure; T1D = Type 1 Diabetes; RAPA = Rapid Assessment of Physical Activity (6--7 indicates active); Dosing Windows and Medications ($n$) reflect operational medication complexity and daily administration density. .}}
\end{table*}

A lead researcher and a sports scientist selected a representative one-month data window for each patient, prioritizing periods with high activity variety and frequency. In addition to continuous wearable streams, clinical background data were collected, including age, gender, tracker type, self-assessed physical activity level (RAPA), height, weight, BMI, primary medical conditions, active medications, recent blood pressure readings, exercise preferences, physical limitations, and lab-measured vs. calculated maximum heart rates. Treatment goals were established by the sports scientist based on individual risk factors and exercise habits. Donor patients were compensated with a Polar Verity Sense sensor (wearable device) for their participation.

\subsection{Participants}
Twenty-six healthcare professionals ($N = 26$) participated in the main study (excluding three pilot participants). Participants had a mean age of 27.0 years (SD = 5.1, range = 22–46); 15 identified as male and 11 as female. All participants were practising Exercise Physiologists, with 7 ($26.9\%$) additionally pursuing a PhD in Human Kinetics. Participants ranged in age from 22 to 46 years ($M = 27.2$, $SD = 5.2$; 15 male, 11 female) and had an average of 4.1 years of experience working in healthcare ($SD = 3.7$, median = 3, range = 1–19).

The sample (Table~\ref{tab:participant_demographics}) demonstrated high domain-specific expertise: 24 participants ($92.3\%$) had direct training or practical clinical experience in cardiac rehabilitation (e.g., patient monitoring and assessment), and 24 ($92.3\%$) routinely worked with patient-generated health data (PGHD), such as accelerometry, heart rate monitoring, RPE scales, or blood pressure tracking. Furthermore, all participants ($100\%$) reported prior usage of generative AI tools (e.g., ChatGPT, Gemini, Claude). Participants' visualization literacy was measured using the MiniVLAT~\cite{pandey_mini-vlat_2023}, which ranges from 0 to 12, with higher scores indicating greater visualization literacy. Participants had a mean MiniVLAT score of 9.15 (SD = 1.49, median = 9.5, range = 6–11). 

\begin{table}[htbp]
  \caption{Participant Demographics and Characteristics ($N = 26$).}
  \label{tab:participant_demographics}
  \begin{tabular}{l r}
    \toprule
    \textbf{Characteristic} & \textbf{Participants} \\
    \midrule
    $N$ & $26$ \\
    Age, mean (SD) & $27.0$ ($5.1$) \\
    Age, range & $22$--$46$ \\
    Gender, female / male & $11$ / $15$ \\
    Healthcare experience, mean years (SD) & $4.1$ ($3.7$) \\
    \midrule
    \multicolumn{2}{l}{\textit{Role}} \\
    \quad Exercise physiologist & $18$ \\
    \quad Exercise physiologist + PhD candidate & $8$ \\
    \midrule
    \multicolumn{2}{l}{\textit{Prior Experience}} \\
    \quad Prior cardiac rehabilitation experience & $24$ \\
    \quad Prior PGHD experience & $24$ \\
    \quad Prior AI/LLM experience & $26$ \\
    \midrule
    \multicolumn{2}{l}{\textit{Visualization Literacy}} \\
    \quad MiniVLAT, mean (SD) & $9.15$ ($1.49$) \\
    \quad MiniVLAT, range & $6$--$11$ \\
    \bottomrule
  \end{tabular}
\end{table}

Participants were randomly assigned across four counterbalanced ordering groups ($G1$--$G4$, $n \approx 6$--$7$ per group) (Table~\ref{tab:study_designOrder}). The study protocol was refined via a pilot study with three sports scientists prior to formal data collection. The study took approximately 90 minutes, and participants were reimbursed with a €50 Amazon gift voucher.


\subsection{AI-Assisted Decision Support Dashboard}
\begin{figure*}[t]
    \centering
    \includegraphics[width=0.6\linewidth]{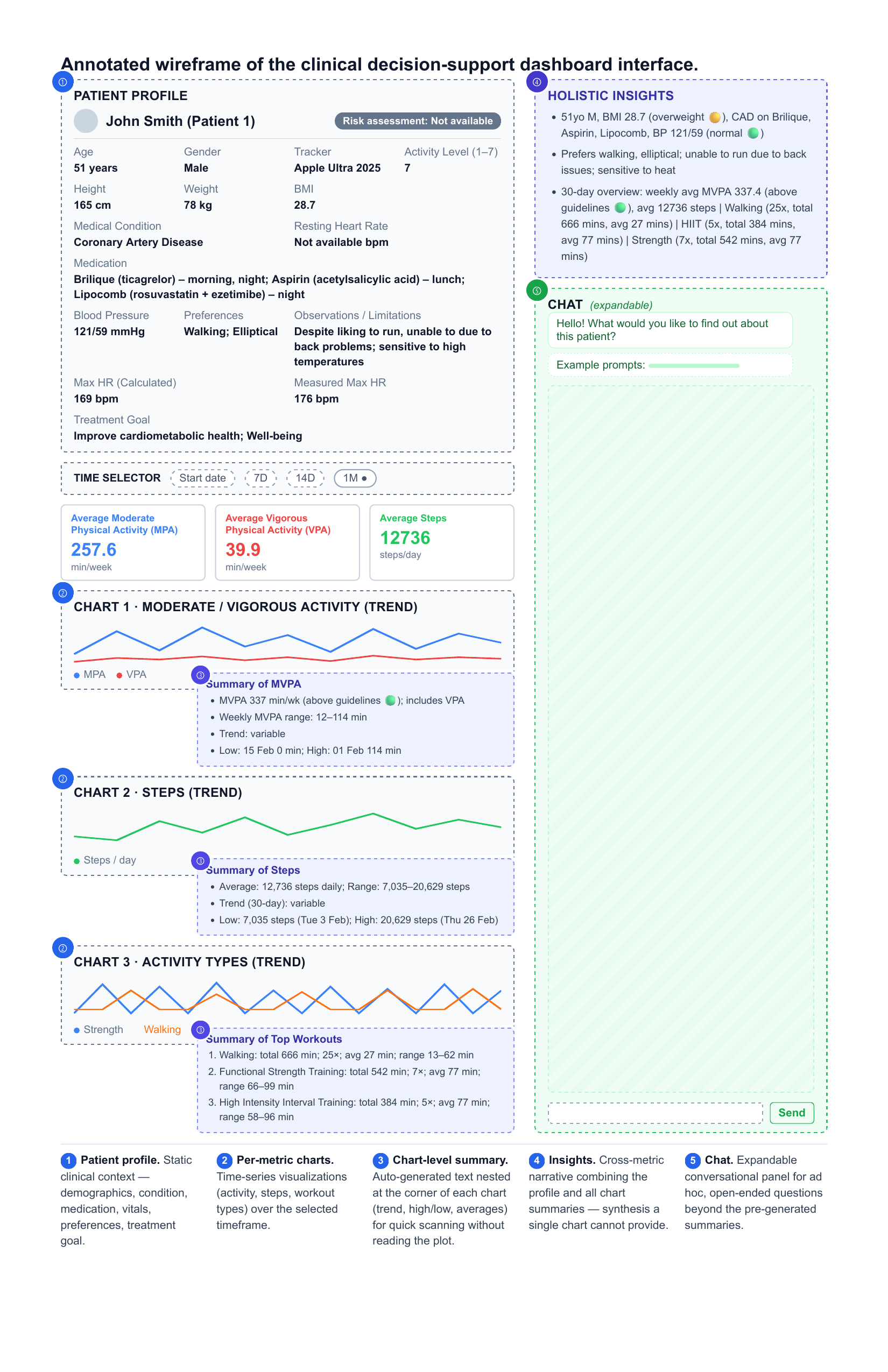}
    \caption{Overview of the clinical decision support system}
    \Description{A five-part dashboard wireframe with annotations labeled one through five across two main columns. On the left side, region one at the top displays a static patient profile containing demographics, health metrics, current medications, blood pressure, calculated and measured maximum heart rate, and treatment goals. Directly below the profile is a time selector with options for seven days, fourteen days, and one month, followed by three summary metric cards displaying average moderate physical activity, average vigorous physical activity, and average steps. Region two contains three stacked line charts over time: Chart one shows moderate and vigorous physical activity trends; Chart two shows daily step counts; Chart three shows activity types, tracking strength training and walking. Region three consists of text summaries embedded at the bottom corner of each chart, providing text key statistics such as ranges, weekly averages, and low or high values. On the right side, region four at the top presents a holistic insights section that synthesizes the patient profile, blood pressure, and thirty-day activity trends into bullet points. Region five, below the insights section, is an expandable, conversational chat panel containing a prompt line and a text input field with a send button for ad hoc questions. At the bottom of the interface, a numbered key details each section: one, Patient profile; two, Per-metric charts; three, Chart-level summary; four, Insights; and five, Chat.}
    \label{fig:wireframe}
\end{figure*}

The AI-CDSS (Figure \ref{fig:wireframe}) was based on the system described in~\cite{pakianathan_exploring_2026,pakianathan_towards_2026} and was further improved by input from exercise physiologists at a cardiac rehabilitation site. The system was developed as a web-based clinical dashboard using Plotly Dash for the front-end interface and n8n to manage the underlying GenAI processing pipeline.

The system aggregated patient information across distinct interface components:
\begin{enumerate}
    \item \textbf{Patient Profile:} Summarized static clinical background, including demographics, medical conditions, current medications, vital signs, physical preferences, and treatment goals.
    \item \textbf{Wearable Data Visualizations and Stream Summaries:} Visualized three continuous tracking streams --- daily step counts, Moderate-to-Vigorous Physical Activity (MVPA) minutes, and activity types. Below each chart, natural language summaries (only for AI condition) provided targeted insights:
    \begin{itemize}
        \item \textit{MVPA Summaries:} Displayed average daily MVPA, alignment with World Health Organization (WHO) physical activity guidelines, parameter ranges, temporal trends, and peak/trough activity days.
        \item \textit{Step Summaries:} Highlighted mean daily steps, range, overall trends over the selected timeframe, and specific days with highest and lowest step counts.
        \item \textit{Activity Summaries:} Itemized individual exercise modalities, specifying total duration, session frequencies, and average session metrics.
    \end{itemize}
    \item \textbf{Holistic Clinical Summary (only for AI condition) \& Risk Stratification:} Located at the top right of the dashboard, the holistic summary presented high-level information (age, BMI, condition, medication, blood pressure), exercise preferences, physical limitations, and overall exercise patterns. A visual traffic-light risk indicator categorized overall patient risk into green (low), yellow (medium), or red (high).
    \item \textbf{Interactive Conversational AI (only for AI condition):} Included a real-time chat interface enabling HCPs to query the information dynamically. We restricted responses related to full exercise plan generation and allowed the chat to generate visualizations such as dynamic charts for enabling data analysis. More information about prompts can be found in the Appendix.
\end{enumerate}

The system was primarily in English but participants could also use the chatbot in their native language. All natural language summaries dynamically refreshed when HCPs modified the visualization timeframe. Visualizations and summary rules were co-designed based on feedback from pre-study iterations, sports scientists, and a senior professor specializing in cardiac rehabilitation. 

To optimize balance between execution latency and analytical capability while responding to user queries, the conversational chat relied on \texttt{gpt-5.4-nano-2026-03-17}, while summary generation utilized \texttt{gpt-4.1-mini}. To prevent numerical hallucination, numerical aggregations (averages, ranges, totals) were pre-calculated programmatically prior to LLM summarization. The dashboard underwent iterative validation across three temporal ranges and four patient cases by three researchers to confirm data accuracy and system robustness.

\subsection{Study Design and Procedure}
\begin{figure}
    \centering
    \includegraphics[width=1\linewidth]{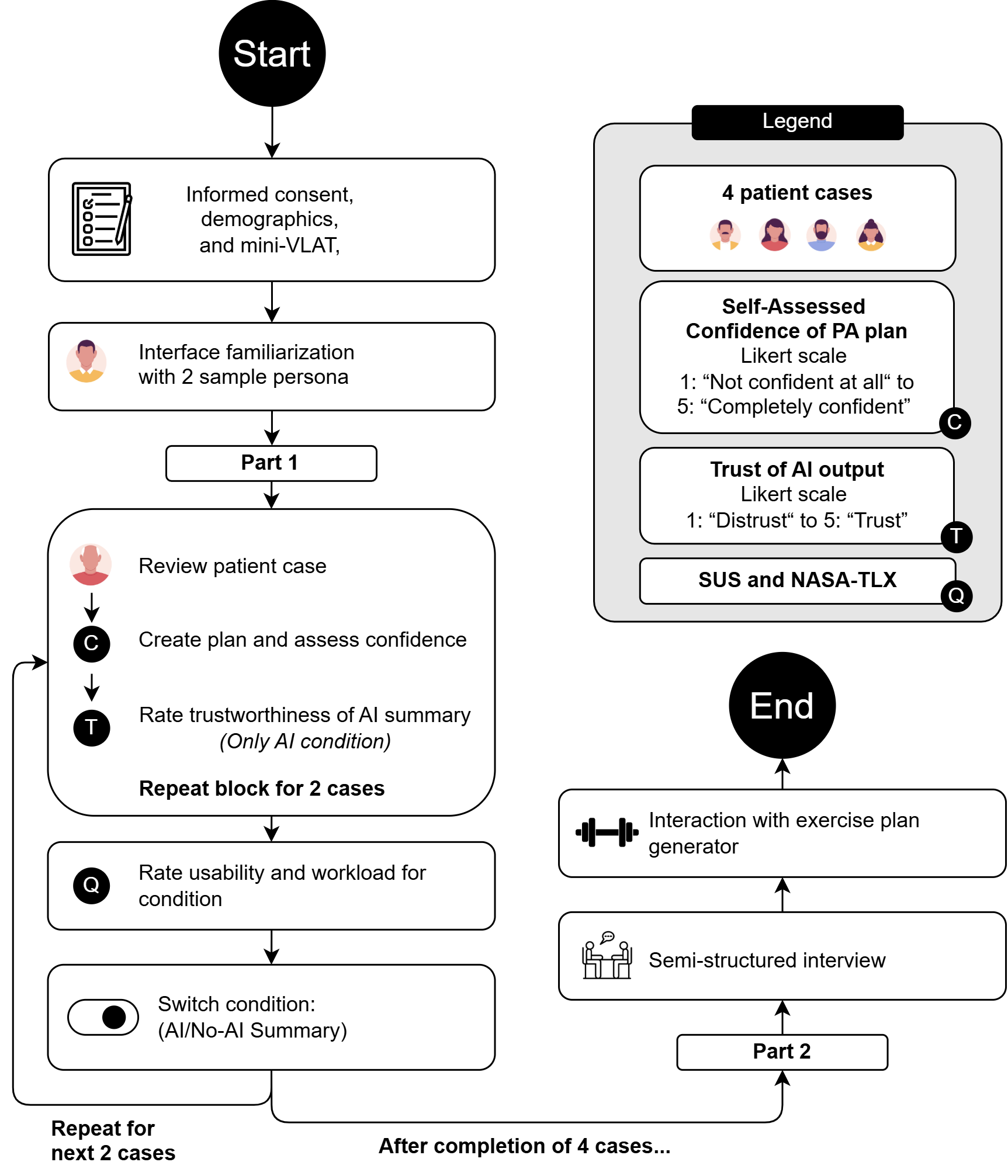}
    \caption{Schematic overview of the study procedure, showing the counterbalanced execution of four patient cases across AI and No-AI conditions (Part 1) followed by the semi-structured interview and interactive hands-on session (Part 2).}
    \Description{A sequential flowchart mapping the study procedure from Start to End across two experimental parts. The main sequence begins at a black circular node labeled Start, pointing downward to Informed consent, demographics, and mini-VLAT, followed by Interface familiarization with two sample personas. The flow enters Part 1, leading into an inner task loop where participants review a patient case, create a physical activity plan while assessing confidence (marked C), and rate the trustworthiness of the AI summary in the AI condition (marked T). A label indicates this block repeats for two cases. After completing two cases, a box labeled Rate usability and workload for condition (marked Q) is reached, followed by a toggle icon box labeled Switch condition: AI or No-AI Summary. An arrow loops back to repeat this sequence for the next two cases. After completing all four cases, the flow transitions to Part 2, leading through a Semi-structured interview, followed by Interaction with exercise plan generator, and terminating at a black circular node labeled End. A legend box on the right explains the circle markers used throughout the flowchart: four patient cases represented by user avatars; letter C for self-assessed confidence of physical activity plan on a one-to-five Likert scale; letter T for trust of AI output on a one-to-five Likert scale; and letter Q for System Usability Scale (SUS) and NASA Task Load Index (NASA-TLX).}
    \label{fig:studyDesignDiagram}
\end{figure}

We employed a within-subjects design in which each HCP created exercise plans for all four patient cases (2 ×  AI-assisted condition, 2 ×  unassisted No-AI baseline condition) (Figure \ref{fig:studyDesignDiagram}). Half of the participants experienced the AI condition first (n = 13) and half experienced the non-AI condition first (n = 13). Condition order was counterbalanced across four groups ($G1$--$G4$) so that AI exposure occurred either during the first two or final two cases, while patient cases were varied across slots using a Latin-square rotation to mitigate ordering, learning, and case-difficulty effects (Table~\ref{tab:study_designOrder}).

The approx. 90 minutes sessions followed a structured procedure:
\begin{enumerate}
    \item \textbf{Onboarding \& Baseline Questionnaire (15 min):} Participants completed an orientation on system features and received instruction on filling out the standardized Excel exercise plan template developed in collaboration with sports scientists. Baseline expectations were recorded.
    \item \textbf{Trial Execution (approx. 45 min):} HCPs reviewed patient profiles and drafted tailored exercise plans for four cases. Participants were encouraged to spend approximately 10 minutes per patient. This timeframe was selected after a pilot session with two exercise physiologists and was intended to create a time-constrained decision-making task rather than reproduce the duration of a complete exercise planning consultation with a patient. In clinical practice, exercise prescription may involve broader assessment of clinical status, medications, exercise capacity, risk factors, and patient preferences and may therefore require substantially more time. Self-reported measures were logged immediately following each individual patient case, as well as upon completing each two-case condition block (AI vs. No-AI).
    \item \textbf{Semi-structured Interview (15 min):} Following plan completion across all four cases, a 15-minute semi-structured  interview was conducted to probe participants' decision-making processes, information preferences, and perceptions of AI support.
    \item \textbf{Hands-on with AI-exercise generator (15 min):} Finally, participants interacted with a high-fidelity early version of an AI physical activity plan generator (Figure \ref{fig:exgen}). Questions were asked about benefits and risks of adopting such a tool and modes of integrating it into their clinical workflows.
\end{enumerate}

\begin{table}[htbp]
\centering
\caption{Counterbalanced Order of Conditions and Patient Cases Across Participant Groups}
\label{tab:study_designOrder}
\small
\setlength{\tabcolsep}{3.5pt} 
\begin{tabular}{lcccc}
\hline
\textbf{Group} & \textbf{Task Slot 1} & \textbf{Task Slot 2} & \textbf{Task Slot 3} & \textbf{Task Slot 4} \\ \hline
\textbf{G1} & AI (Case 1)    & AI (Case 2)    & No-AI (Case 3) & No-AI (Case 4) \\
\textbf{G2} & No-AI (Case 2) & No-AI (Case 1) & AI (Case 4)    & AI (Case 3)    \\
\textbf{G3} & AI (Case 3)    & AI (Case 4)    & No-AI (Case 1) & No-AI (Case 2) \\
\textbf{G4} & No-AI (Case 4) & No-AI (Case 3) & AI (Case 2)    & AI (Case 1)    \\ \hline
\end{tabular}
\end{table}

\subsection{Measures}
All measures were translated to the native language of the participants.

\subsubsection{Pre-Task Expectation Measures}
At the end of the orientation, prior to task exposure, baseline expectations regarding the AI decision-support system were captured via the \textit{iExpect} questionnaire ~\cite{villa_inventory_2026}. The scale assesses three distinct subscales: Expected Usefulness (EU), Expected Usability (US), and Affective Evaluation (AE). An overall composite expectation score was calculated as the mean across all items, with two negatively phrased items reverse-scored.

\subsubsection{Per-Scenario Measures (4 ×  per participant)}
Immediately after finalizing an exercise plan for each patient scenario, the following metrics were logged:
\begin{itemize}
    \item \textbf{Plan Confidence:} HCPs rated their confidence in the clinical appropriateness of their generated exercise plan on a 5-point Likert scale (1 = Very Low, 5 = Very High).
    \item \textbf{Trust in AI Output:} Collected exclusively in the AI condition, HCPs rated their agreement with single-item statements regarding their trust in the provided AI summaries and chat responses on a 5-point Likert scale (1 = Distrust, 5 = Trust). We did not use a validated scale, and used the phrasing: ``Trust in AI outputs''

\end{itemize}

\subsubsection{Per-Condition Measures (2 ×  per participant)}
Following the completion of each condition block (AI vs. No-AI), participants completed two standardized instruments:
\begin{itemize}
    \item \textbf{System Usability Scale (SUS)~\cite{brooke_sus-quick_1996}:} A 10-item questionnaire yielding a composite usability score ranging from 0 to 100, where higher scores indicate greater perceived usability.
    \item \textbf{NASA Task Load Index (NASA-TLX)~\cite{hart_development_1988}:} Subjective mental workload was measured across six standard dimensions (Mental Demand, Physical Demand, Temporal Demand, Performance, Effort, and Frustration). Scores were calculated as an unweighted average across subscales, where lower scores indicate lower overall  workload.
\end{itemize}

\subsubsection{Expert Plan Quality Evaluation}
To evaluate the objective clinical quality of the 104 generated exercise plans (26 participants ×  4 cases), each plan was independently evaluated by two clinical experts drawn from a pool of six expert raters with nine to 20 years of experience using an uncrossed review assignment (Table \ref{tab:expert_background}). Experts evaluated plans across three primary dimensions on a 5-point Likert scale: Q1 (\textit{Clinical appropriateness and Safety}, 1 =  Unsafe and Not Appropriate, 5 = Safe and Appropriate), Q2 (\textit{Personalization}, 1 = Not personalized, 5 = Personalized), and Q3 (\textit{Plan Clarity}, 1 = Unclear, 5 = Clear). The six experts took approximately 200 minutes to review the cases and the exercise plans created by the study participants and were reimbursed with a €200  Amazon voucher.

\begin{table}[t]
\centering
\caption{Background and cardiac rehabilitation experience of the expert evaluators.}
\label{tab:expert_background}
\begin{tabular}{p{0.08\linewidth} p{0.62\linewidth} p{0.20\linewidth}}
\toprule
\textbf{Expert} & \textbf{Professional background} & \textbf{CR experience} \\
\midrule
E1 & Exercise physiologist, researcher, and university lecturer in clinical exercise and cardiac rehabilitation & 14 years \\
E2 & Exercise physiologist, professor, and researcher in exercise physiology, sports medicine, and rehabilitation & 10 years \\
E3 & Exercise physiologist and assistant professor & 14 years \\
E4 & Exercise physiologist and university professor & 20 years \\
E5 & Exercise physiologist, researcher, and PhD candidate in cardiovascular rehabilitation & 10 years \\
E6 & Exercise physiologist working in cardiac rehabilitation & 9 years\\
\bottomrule
\end{tabular}
\end{table}

\subsection{Data Analysis}

\paragraph{Quantitative Data}
We used linear mixed-effects models to compare the AI-supported and No-AI conditions while accounting for repeated observations within participants. Plan confidence was analyzed at the patient-case level, while NASA-TLX and SUS were analyzed at the condition-block level. Expert-rated plan quality was analyzed using mixed-effects models with random intercepts for participant, rater, and plan. We additionally examined whether visualization literacy moderated the effect of AI support by including a Condition × MiniVLAT interaction. MiniVLAT was mean-centered for analysis. We report estimated coefficients (\(\beta\)), 95\% confidence intervals, and \(p\)-values. Descriptive statistics were used to summarize participant characteristics, questionnaire measures, and conversational AI usage. 

\paragraph{Qualitative Data}
All interviews were conducted in [blinded] and then transcribed using OpenAI whisper and verified by single author. Subsequently the transcript was translated to English using ChatGPT 5.4 and verified by the same researcher. The transcripts were analyzed using qualitative content analysis by Mayring \cite{mayring_qualitative_2014}, a structured approach to systematically categorize qualitative data. The initial categories were discussed among the first two authors, and fourth authors were iteratively refined during analysis to accommodate patterns emerging from the data. Similarly, the chat transcripts were analysed using the same method. Separate coding schemes were developed for the interview transcripts and conversational AI interaction logs.

Chat queries were categorized using the qualitative coding scheme and summarized by category and participant characteristics. Interview data were analyzed using content, following an iterative process of coding, grouping related codes, and developing themes that characterized how and under what circumstances HCPs used AI support.

\paragraph{Ethics}
This study was conducted in accordance with the Declaration of Helsinki and GDPR. Ethical approval for the collection, pseudonymization, and secondary use of patient-generated health data was granted by the Ethics Committee of [Blinded]. User Study: Ethical approval for the user evaluation study with healthcare professionals interacting with the system was granted by the Ethics Committee of [Blinded]. Informed consent was obtained from all participating patients and healthcare professionals in their respective local languages prior to data collection. 
\section{Findings}

We first report how AI support related to physical activity planning outcomes and how these effects varied across HCPs (RQ1). We then examine how and when HCPs used summaries, visualizations, and conversational AI during sensemaking and planning (RQ2), drawing on interviews and interaction logs. Finally, we report findings from the exploratory AI plan-generation probe.
\subsection{AI Support Showed Heterogeneous Effects on Physical Activity Planning}

\subsubsection{Planning Outcomes Were Similar on Average Across Conditions}

Across the full sample, we found no significant overall differences in planning outcomes between the AI-supported and No-AI conditions. For the expert-rated plans, inter-rater reliability was modest, with Krippendorff's $\alpha=.26$ for appropriateness and safety, $\alpha=.12$ for personalization, and $\alpha=.40$ for clarity. Despite the modest reliability, ratings were within one scale point for 90.4\%, 81.7\%, and 89.4\% of plans, respectively. Mixed-effects models showed small, non-significant differences between conditions for appropriateness and safety ($\beta=-.096$, 95\% CI [-.316, .124], $p=.392$), personalization ($\beta=-.096$, 95\% CI [-.297, .104], $p=.348$), and clarity ($\beta=.058$, 95\% CI [-.124, .239], $p=.534$).

Plan confidence was slightly higher in the AI-supported condition ($M=3.87$, $SD=.97$) than in the No-AI condition ($M=3.69$, $SD=.96$), although not significant ($\beta=.17$, 95\% CI [-.05, .39], $p=.126$). Perceived workload was also slightly lower with AI support (NASA-TLX: $M=51.19$, $SD=17.64$) than without AI support ($M=52.82$, $SD=17.28$), but this difference was not significant ($\beta=-1.63$, 95\% CI [-4.87, 1.60], $p=.321$). Usability was rated highly in both conditions, with SUS scores of $M=78.56$ ($SD=12.96$) for the AI-supported interface and $M=77.40$ ($SD=13.92$) for the No-AI interface; the difference was not significant ($\beta=1.15$, 95\% CI [-1.58, 3.89], $p=.408$). Participants reported relatively high trust in AI outputs (\(M=3.90, SD=.77\)), with little within-participant variation. Trust was not significantly associated with visualization literacy, plan confidence, or any expert-rated plan-quality measure (all \(p>.20\)). We next examined whether the relative effect of AI varied across HCPs.

\subsubsection{Visualization Literacy Moderated AI Effects on Plan Quality and Workload}

Participants had a mean MiniVLAT score of $9.15$ ($SD=1.49$, range = 6--11). MiniVLAT scores were reverse-centered around the sample mean such that higher values represented lower visualization literacy.

Visualization literacy moderated the relative effect of AI support on expert-rated plan quality (Table~\ref{tab:visualization_literacy_effects}). The interaction was significant for aggregate plan quality ($\beta=.13$, $p=.007$), as well as clinical appropriateness and safety ($\beta=.18$, $p=.008$), with AI becoming relatively more favorable as visualization literacy decreased. Moderation effects for personalization, plan clarity, and plan confidence were not statistically significant.

Visualization literacy also moderated perceived workload ($\beta=3.13$, $p=.001$). Although average NASA-TLX scores were similar between the AI-supported ($M=51.19$, $SD=17.64$) and No-AI conditions ($M=52.82$, $SD=17.28$), the relative workload advantage of AI increased with higher visualization literacy. Together, these findings show different patterns across outcomes: AI became relatively more favorable for plan quality at lower visualization literacy, while its relative workload advantage increased at higher visualization literacy.

\begin{table*}[t]
  \centering
  \caption{Interaction effects of condition and visualization literacy.}
  \label{tab:visualization_literacy_effects}
  \begin{tabular}{l r c r}
    \toprule
    \textbf{Outcome} & \textbf{Condition $\times$ Lower visualization literacy $\beta$} & \textbf{95\% CI} & \textbf{$p$} \\
    \midrule
    Confidence & $-0.10$ & $-0.25$ to $0.06$ & $.223$ \\
    Clinical appropriateness/safety & $+0.18$ & $0.05$ to $0.32$ & $.008$ \\
    Personalization & $+0.13$ & $-0.01$ to $0.27$ & $.070$ \\
    Plan clarity & $+0.07$ & $-0.02$ to $0.17$ & $.130$ \\
    Aggregate quality & $+0.13$ & $0.04$ to $0.23$ & $.007$ \\
    NASA-TLX & $+3.13$ & $1.26$ to $5.01$ & $0.001$ \\
    \bottomrule
  \end{tabular}
\end{table*}

\subsection{HCPs Used AI Selectively Across Data Sensemaking and Physical Activity Planning}

Across the AI-supported cases, participants submitted 152 queries to the conversational interface, while six participants (23\%) did not use the chat at all. We categorized these queries into four types according to the information or support requested (Table~\ref{tab:chat_categories}). We integrate these interaction-log results with the interview findings below to characterize how AI support was used across different parts of data sensemaking and physical activity planning. Representative chat requests are shown in Table~\ref{tab:representative_chat_queries}.

\begin{table*}[t]
\centering
\caption{Categorization and frequency of HCP chat queries ($N=152$ total queries).}
\label{tab:chat_categories}
\small
\begin{tabularx}{\textwidth}{X r}
\toprule
\textbf{Query Category} & \textbf{Count ($n$)} \\
\midrule
Trends and patterns in physical activity behaviour & 60 \\
Information about patient & 8 \\
Medication side effects and interaction with physical activity & 37 \\
Physical activity recommendations and medication-related safety measures & 47 \\
\midrule
\textbf{Total} & \textbf{152} \\
\bottomrule
\end{tabularx}
\end{table*}

\begin{table*}[t]
\caption{Representative chat queries translated into English and grouped by request type.}
\label{tab:representative_chat_queries}
\small
\begin{tabularx}{\textwidth}{p{3.4cm} p{1.0cm} X}
\toprule
\textbf{Request type} & \textbf{ID} & \textbf{Representative chat request} \\
\midrule

\multirow{4}{3.4cm}{\textbf{Trends and patterns in physical activity behaviour}}
& P4  & ``When was the last time she had moments of vigorous exercise?'' \\[2pt]
& P21 & ``How regularly did the person do HIIT this month?'' \\[2pt]
& P23 & ``In a normal month, on which day does he do the most cycling?'' \\[2pt]
& P9  & ``What are the activities and on which days does the person usually do them?'' \\

\midrule

\multirow{3}{3.4cm}{\textbf{Medication side effects and interaction with physical activity}}
& P12 & ``Is any special attention needed regarding the medication and training?'' \\[2pt]
& P7  & ``Give me a summary of the medication the person takes and the respective precautions to consider for exercise prescription.'' \\[2pt]
& P3  & ``What are aspirin and atorvastatin for?'' \\

\midrule

\multirow{7}{3.4cm}{\textbf{Physical activity recommendations and related safety measures}}
& P10 & ``Should the strength focus prioritize lower-extremity and core endurance to support sustained cycling mechanics?'' \\[2pt]
& P10 & ``Should strength sessions be capped at around 30--45 min, focusing on lower extremities and core?'' \\[2pt]
& P8 & ``Now for session 2, let's introduce running with some low-impact plyometric exercises, then add the elliptical.'' \\[2pt]
& P6 & ``With BP 138/86 and use of losartan/metformin, what is the intensity target (RPE 11--13 or 12--14) and the rest time between sets to minimize blood pressure spikes?'' \\[2pt]
& P20 & ``Should the target intensity be kept at 60\% of HR reserve for walking/elliptical and days with more intense intervals be limited due to tolerance risk (CAD + ticagrelor/aspirin)?'' \\

\midrule

\multirow{3}{3.4cm}{\textbf{Information about patient}}
& P23 & ``Characterize this person's day-to-day life.'' \\[2pt]
& P7  & ``And what limitations do you observe that are useful for the prescription?'' \\

\bottomrule
\end{tabularx}
\end{table*}

\subsubsection{AI Supported Rapid Overview and Extraction of Longitudinal Activity Patterns}

Trends and patterns in physical activity were the most frequent type of conversational request, accounting for 60 of the 152 queries. A further eight queries concerned general patient information. The interaction logs were consistent with interview accounts showing that participants used both AI-generated summaries and conversational AI to access information that could otherwise be derived manually from the visualizations.

\paragraph{Summaries Provided Quick Overviews and Supported Validation}

Participants commonly described the AI-generated summaries as a faster route to information they would otherwise derive manually from the graphs. P10 explained that the system reduced the need to work ``day by day, graph by graph,'' while P5 valued having the patient's activity already summarized: ``I look here [at the holistic summary] and I already know everything the person does.'' P6 similarly emphasized the reduced calculation effort, particularly under time pressure.

For some participants, however, the summaries were not a substitute for inspecting the underlying data. P1 described the holistic summary as ``a validation of the information'' they had already worked out mentally. P3 preferred to ``look at the graphs and try to understand the overall picture'' and consult the summaries only when uncertain, while P7 similarly preferred going ``chart by chart.'' P16 described becoming ``more confident after it said that ... what I wanted to do was fine,'' illustrating how AI support could also be used to confirm an intended decision.

Thus, participants used summaries both to obtain information more quickly and to check interpretations they had developed from the visualizations.

\paragraph{Holistic Summaries Oriented Attention During Sensemaking}

The holistic summary also helped some HCPs establish an initial overview of the patient and identify where further inspection was needed. P21 described it as immediately answering the question, ``okay, who do I have here?'' Visual status cues within the summaries also directed attention. P6 noted that the traffic-light display ``immediately draws the eye,'' while P22 used the red indicators to identify information to prioritize during subsequent inspection.

These cues therefore influenced what some participants attended to first. However, this approach was not universal. P7 preferred reconstructing the patient overview directly from the visualizations: ``I prefer going chart by chart.'' P14, by contrast, used the holistic summary to understand whether patients were meeting physical activity guidelines and to orient the direction of the prescription.

\paragraph{Chat Supported Targeted Extraction of Activity Patterns}

Conversational AI provided a more targeted way of retrieving information from longitudinal activity data. P9 described asking ``what the person did and on which days, so that I did not have to look at the graph.'' P11 queried weekly exercise frequency and whether a particular day was typically a rest day, while P23 used chat to examine whether weekly physical activity habits formed a broader monthly pattern.

These interactions show how participants used natural-language queries to retrieve specific information from longitudinal data without manually inspecting each visualization.

\subsubsection{HCPs Turned to AI for Medication and Safety Questions Under Uncertainty}

Medication side effects and interactions with physical activity accounted for 37 of the 152 chat queries. Participants used the conversational interface to ask about medication purpose, potential effects on exercise, and precautions relevant to physical activity planning. P3 described asking ``what each medication was ... and whether it would influence physical activity.'' P12 similarly asked whether a patient's medication required ``special attention'' during physical activity training. P21 used the chatbot to check whether metformin increased hypoglycemia risk and reported feeling reassured when the response clarified the relationship.

Participants' confidence-related accounts were also particularly connected to medication and exercise interactions. P1 described medication-related queries as useful for understanding ``something that I do not master,'' and later said that AI support ``gave me more confidence'' that the training plan was appropriate. P3 similarly reported being ``more confident in the ones I created with AI, again because of the medication.''

The importance of this support became particularly apparent when a case exposed information that participants did not feel able to resolve readily from the dashboard and their own knowledge. P4 explained that Patient 2 was ``taking several medications'' whose effects they did not know and that this information was needed ``to make a better prescription.'' P24 similarly reported that because Patient 2 had ``a lot of medication,'' they asked the chat immediately rather than spending additional time searching elsewhere. Patient 2 also had diabetes and a comparatively medication-heavy profile (Table~\ref{tab:patient_demographics}). P4 described the additional support as giving them ``a bit more confidence in what I was prescribing.'' This pattern was also consistent with the patient-level confidence analysis, in which confidence was higher with AI support for Patient 2.

The perceived need for such support was not shared by all participants. P25 explained, ``I know what the medications are. I know what implications they have for practice.'' P22 similarly reported being ``comfortable with all the information'' and with their knowledge of the relevant conditions, while acknowledging that AI could be useful to colleagues with less knowledge or experience. These accounts suggest that medication-related AI support became particularly salient when HCPs perceived a gap between the information required for planning and the knowledge readily available to them.

\subsubsection{AI Use Extended from Information Seeking to Physical Activity Recommendations}

Conversational use also extended beyond interpreting patient information toward determining how that information could inform a physical activity plan. Physical activity recommendations and medication-related safety considerations accounted for 47 of the 152 chat queries, making this the second most frequent category after activity-pattern queries.

Participants asked about exercise intensity, training structure, exercise type, and safety considerations. P8 asked the system to suggest an interval-training session and reported that it returned effort and recovery periods and heart-rate zones. Other requests concerned session intensity, strength-training duration, exercise type, and recommendations aligned with activities patients were already performing. P20 used the conversational interface to calculate a target based on ``70\% of maximum heart rate.''

Participants also used existing physical activity behaviour to inform patient-specific planning. P5 noted that repeated walking activity ``gives you an idea that the person really likes walking,'' while P4 described using information about when and how a patient typically trained to support the planning process. P12 perceived this type of support as potentially enabling ``more individualized or more accurate'' decisions.

We treat this as a participant perception rather than evidence that AI objectively improved personalization, as expert ratings did not show a significant overall condition effect on personalization and showed substantial disagreement over what constituted personalization and adequate progression.

Some participants imagined extending conversational interaction further toward direct manipulation of a developing plan. P6 proposed being able to tell the system to ``remove the Monday walk and add gym instead,'' after which the HCP would ``just correct it.'' These interactions show how conversational AI use could move from asking what was occurring in the patient data toward asking how that information should inform an exercise prescription.

\subsubsection{AI Engagement Was Selective Across HCPs}

The availability of AI did not produce a single AI-supported workflow. Six of the 26 participants did not use the conversational interface at all, while others used it only for particular questions or cases. Participants selected between summaries, chat, and visualizations depending on what they were trying to understand and whether they expected AI to provide additional value.

P21 described returning to AI when the graphs became confusing: ``I would look down and immediately clear up the doubts.'' P12 similarly regarded chat as a resource to consult whenever uncertainty arose. Others deliberately maintained a graph-first workflow. P2 and P22 preferred inspecting the visualizations directly and reported confidence in interpreting the charts and understanding the medication information. P14, by contrast, used the holistic summary to determine whether patients were meeting activity guidelines and to orient the direction of the prescription. P19 suggested that experienced professionals could interpret the graphs directly, whereas summaries might be especially useful to someone ``just starting out.''

These differences indicate selective appropriation of AI support based on perceived expertise, uncertainty, trust, AI acceptance, resource constraints, and the demands of the particular case.

\paragraph{Chat Use Varied Descriptively by Visualization Literacy and Experience}

For descriptive comparison, we examined the distribution of chat request types by visualization literacy and professional experience. Lower and higher visualization literacy were defined using a MiniVLAT split at 9.5 ($n=13$ per group), while lower and higher professional experience were defined as $\leq 3$ years ($n=17$) and $>3$ years ($n=9$), respectively.

\begin{figure}
    \centering
    \includegraphics[width=1\linewidth]{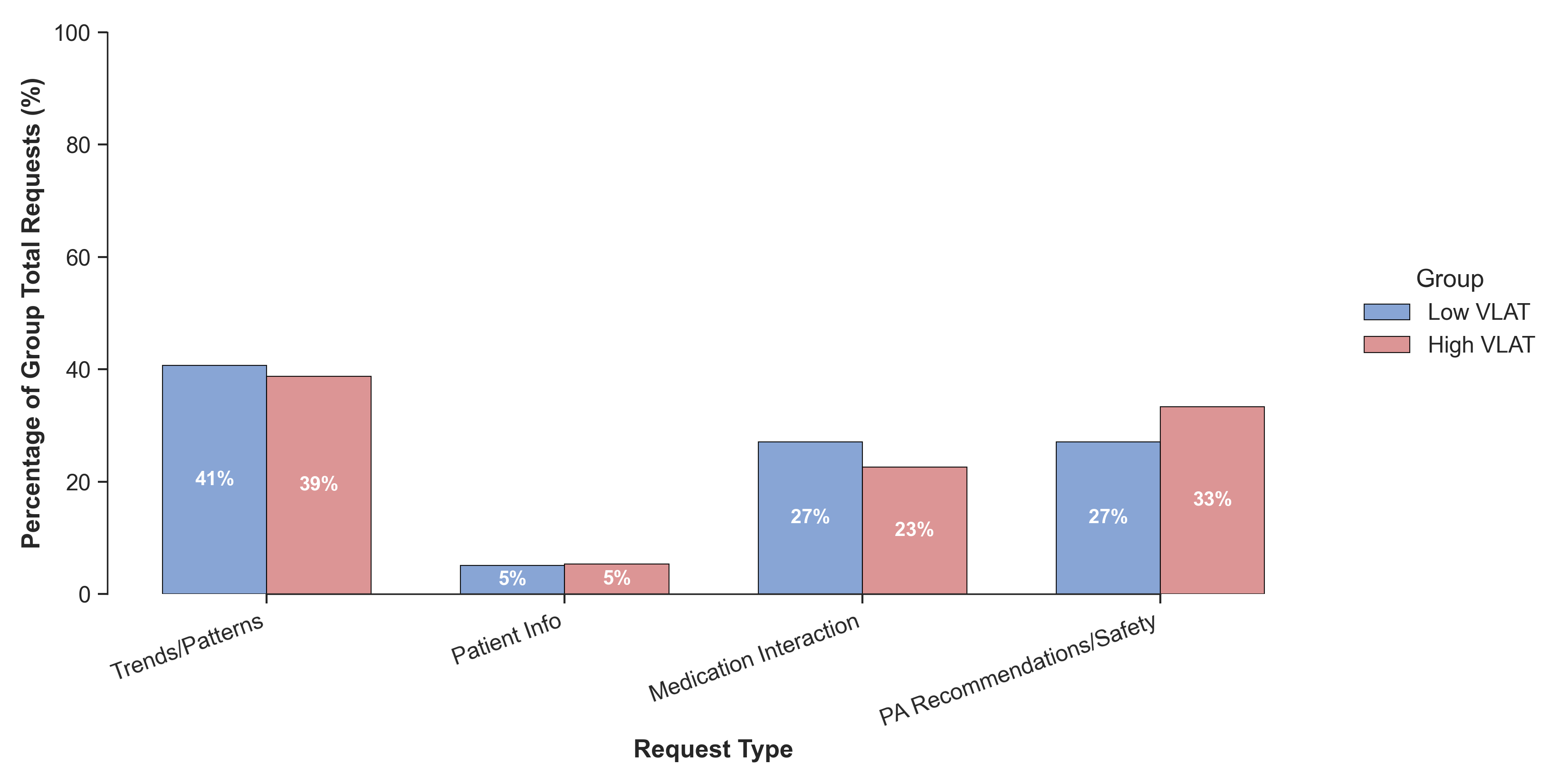}
    \caption{Distribution of chat request types grouped by visualization literacy.}
    \Description{
    Grouped bar chart showing the distribution of chat request types for participants with lower and higher visualization literacy. Among the lower-VLAT group, 41\% of queries concerned trends and patterns, 5\% patient information, 27\% medication interactions, and 27\% physical activity recommendations or safety. Among the higher-VLAT group, the corresponding percentages were 39\%, 5\%, 23\%, and 33\%, respectively.}
    \label{fig:totalchatVlat}
\end{figure}

Request distributions were broadly similar across visualization-literacy groups (Figure~\ref{fig:totalchatVlat}). Trends and patterns accounted for the largest share of queries in both groups (41\% for lower-VLAT participants and 39\% for higher-VLAT participants). Lower-VLAT participants made a slightly larger proportion of medication-related queries (27\% vs.\ 23\%), whereas higher-VLAT participants made a somewhat larger proportion of physical activity recommendation and safety-related queries (33\% vs.\ 27\%).

\begin{figure}
    \centering
    \includegraphics[width=1\linewidth]{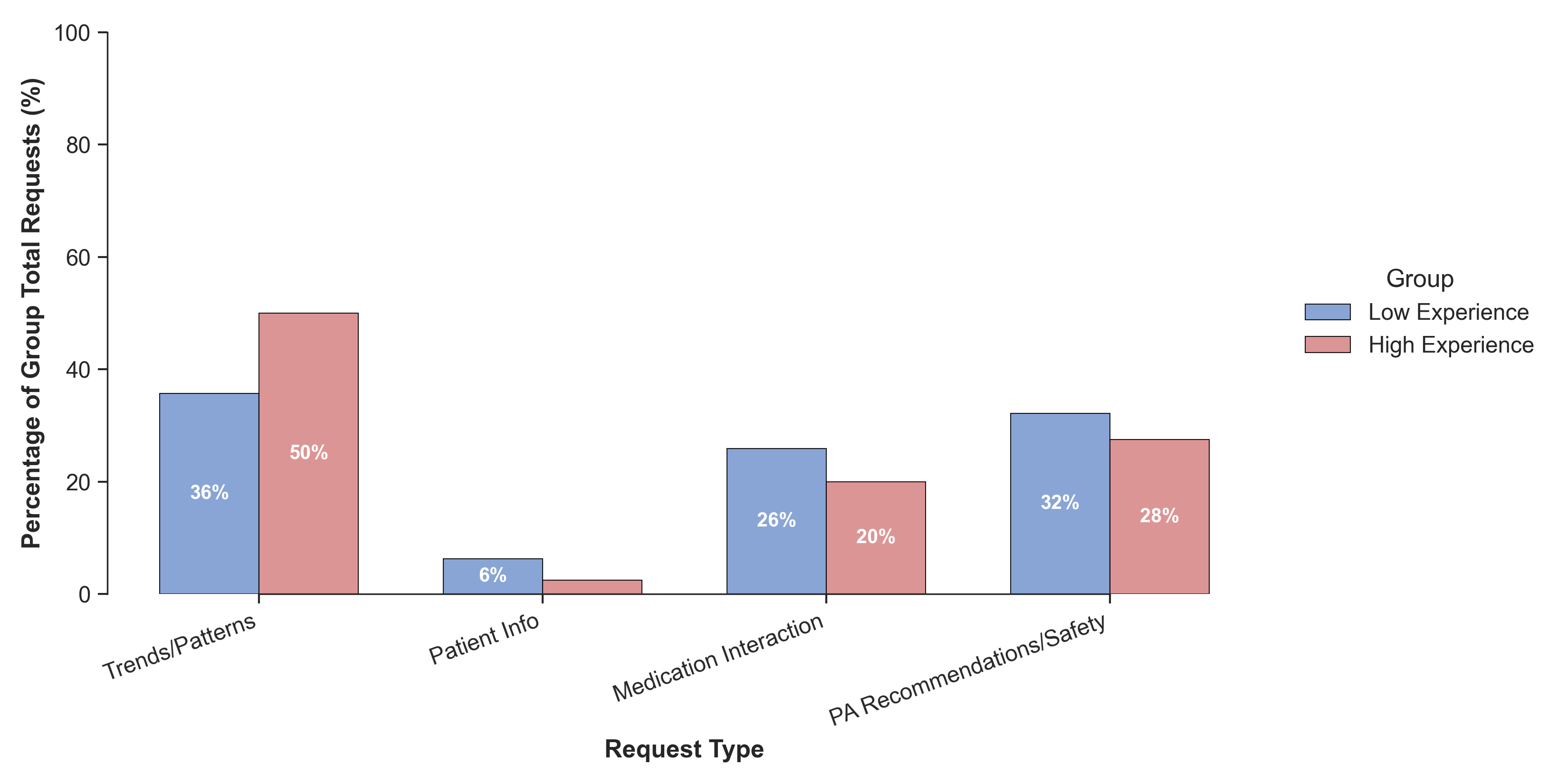}
    \caption{Distribution of chat request types grouped by professional experience.}
    \Description{
    Grouped bar chart showing the distribution of chat request types for participants with lower and higher professional experience. Among the lower-experience group, 36\% of queries concerned trends and patterns, 6\% patient information, 26\% medication interactions, and 32\% physical activity recommendations or safety. Among the higher-experience group, the corresponding percentages were 50\%, 2\%, 20\%, and 28\%, respectively.}
    \label{fig:totalchatExp}
\end{figure}

Descriptive differences were somewhat larger across experience groups (Figure~\ref{fig:totalchatExp}). Trends and patterns accounted for 50\% of queries among higher-experience participants compared with 36\% among lower-experience participants. Medication-related queries (26\% vs.\ 20\%) and recommendation or safety-related queries (32\% vs.\ 28\%) represented somewhat larger shares among lower-experience participants.

We treat these comparisons descriptively; the groupings characterize patterns of chat use rather than establish differences between participant groups.

\paragraph{Non-use Could Reflect Existing Knowledge or Preferred Workflows}

Non-use of AI was also deliberate for some participants. P25 stated, ``I know what the medications are. I know what implications they have for practice.'' P22 similarly said they were ``comfortable with all the information'' and their knowledge of the relevant pathologies, while acknowledging the potential value of AI support for colleagues with less knowledge and experience. P19 likewise suggested that experienced professionals could interpret the graphs directly, whereas summaries might be particularly useful to someone ``just starting out.''

A minority preferred their own interpretation of the data with little or no AI support. P13 and P2 drew conclusions primarily from the graphs, P3 used summaries only when uncertain, and P2 mainly attended to the visualizations and referred to the summaries only if their ``eyes happen to catch [the summary].''

\paragraph{Verification and Trust Shaped AI Engagement}

Participants also expressed concerns about the reliability and broader implications of AI support. P9 noted that AI output ``can also be wrong,'' while acknowledging that their own decision had nevertheless been influenced by AI. P9 further worried that over-reliance could reduce HCPs' motivation to independently ``research ... read and check [the training plan it recommends]'' and could remove the ``emotional component'' from the exercise prescription process.

P2 similarly preferred analyzing some patterns independently and argued that useful AI responses depended on providing sufficiently detailed patient context. P20 used the chat to ``confirm doubts'' about medication and exercise structure. Together with P16's use of AI to confirm that an intended decision was appropriate, these accounts show that AI could be used as a source of confirmation while concerns about its reliability continued to shape engagement.

\paragraph{Conversational Support Could Introduce Interaction Costs}

Although summaries were often perceived as time-saving, conversational interaction could introduce additional work. P13 felt that formulating a question through the chat interface could ``take too long.'' The availability of multiple information sources also created additional choices about where to seek information.

P6 described uncertainty about whether to use summaries or chat and worried that chat might sacrifice precision, concluding that they would retain the summaries but potentially remove the conversational component. P7 instead primarily inspected the charts and used chat only when specific questions arose. Thus, adding AI support did not simply reduce effort; it also required participants to decide which source to consult and when AI support was worth invoking.

Overall, participants' engagement reflected an assessment of whether AI offered additional value relative to their existing knowledge and expertise, the demands of the case, and the time and effort required to consult it.

\subsection{HCPs Valued AI-Generated Plans as Editable Starting Points While Retaining Final Control}

Following the main study tasks, participants interacted with a separate prototype AI exercise-plan generator (Figure~\ref{fig:exgen}). We treat these findings as exploratory and separate from the AI-supported versus No-AI comparison reported above.

Participants saw potential value in AI generating an initial plan that could reduce planning work and time and improve efficiency. At the same time, they raised concerns about loss of professional autonomy and the risk of over-reliance.

Several participants explicitly preferred a workflow in which AI produced the initial draft while the HCP retained control over the final prescription. P6 described the generated plan as ``a baseline for the [healthcare] professional to then work from.'' P3 similarly described the workflow as, ``We look at what has already been generated ... and edit what is already there.'' P23 suggested that the generator could ``give you a set of activities that you can accept or reject, and then you make the plan.''

Participants also raised concerns about where decision authority should sit, how easily generated plans could be inspected and edited, and whether automation might gradually replace active professional judgment. P2 drew this boundary explicitly: ``The decision must always come from the professional.'' P23 similarly warned that repeated reliance could progressively weaken verification: ``A day will come when you will no longer even check and you will trust it 100\%.''

Overall, the generator highlighted a tension between reducing planning effort through AI assistance and maintaining deliberative decision making, active verification, and professional judgment in exercise planning.

\section{Discussion}



\begin{figure*}[t]
    \centering
    \includegraphics[width=1\linewidth]{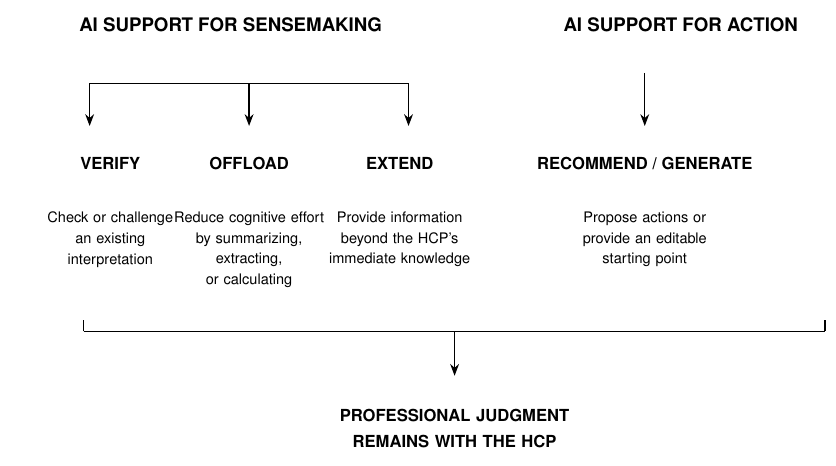}
    \caption{Conceptualization of AI support across professional sensemaking and action in personalized physical activity planning based on our study finding. AI may verify interpretations, offload information-processing work, extend professional knowledge, or support action through recommendations and plan generation, while final judgment remains with the HCP.}
    \Description{A conceptual framework diagram divided into two primary categories at the top: Artificial Intelligence Support for Sensemaking on the left, and Artificial Intelligence Support for Action on the right. Under Artificial Intelligence Support for Sensemaking, three arrows point down to distinct subcategories: One, Verify, defined as checking or challenging an existing interpretation; Two, Offload, defined as reducing cognitive effort by summarizing, extracting, or calculating; Three, Extend, defined as providing information beyond the healthcare provider's immediate knowledge. Under Artificial Intelligence Support for Action, a single arrow points down to a subcategory labeled Recommend or Generate, defined as proposing actions or providing an editable starting point. A bracket connects all four subcategories across both sides to a single downward-pointing arrow at the bottom, which terminates at a central outcome statement: Professional Judgment Remains with the Healthcare Provider.}
    \label{fig:aiFramework}
\end{figure*}

Our study examined how AI support becomes integrated into the process of practising HCPs interpreting real-world longitudinal patient data and constructing personalized physical activity plans. Overall, AI support did not uniformly improve workload, usability, confidence, or plan quality. The relative benefit for plan quality was greater among participants with lower visualization literacy  but its workload benefit increased at higher visualization literacy. The value of AI support emerged selectively amongst the participants within the planning process (Figure \ref{fig:aiFramework}). HCPs used AI to \textit{verify interpretations}, \textit{offload information-processing tasks}, and \textit{extend their knowledge} when encountering uncertainty. Taken together, these findings suggest that the value of AI may lie more in complementing professional capabilities in decision-making rather than universally improving outcomes.

\subsection{AI as Selective Complement to Professional Capabilities}
\subsubsection{AI Effects Varied with Visualization Literacy}

Reflecting on RQ 1, our results suggest that the value of AI support was uneven across HCPs. Visualization literacy significantly moderated the effect of AI support on both expert-rated plan quality and perceived workload (Table \ref{tab:visualization_literacy_effects}). Regarding plan quality, AI became more favourable as visualization literacy decreased including aggregate plan quality and and clinical appropriateness and safety. This suggests that AI support maybe be valuable were the sensemaking task places greater demand on capabilities that varies across the HCPs. Additionally, visualization literacy also moderated perceived workload, where the relative workload advantage of AI increased with higher visualization literacy. Thus AI support benefitted professionals differently; lower visualization literacy was associated with a more favourable relative effect on plan quality, whereas higher visualization literacy was associated with a greater relative workload advantage. 

These findings aligns with prior work showing that expertise and self-reliance can shape engagement with AI-assisted decision support~\cite{rajashekar_human-algorithmic_2024,kuper_psychological_2025} and that AI integration should consider end-user expertise for clinical usefulness~\cite{zajac_it_2024}. Küper et al.~\cite{kuper_psychological_2025}, for example, argue that AI systems should support informed reliance among less experienced clinicians while enabling experienced clinicians to critically engage with, rather than simply defer to, AI recommendations. This recommendation is also reflected in Zhu et al.'s study, where AI provided cognitive scaffolding for junior clinicians while supporting adversarial verification among more experienced clinicians~\cite{zhu_augmenting_2026}.

Taken together, this emphasises the need for considering the fit between the professionals capabilities, task demands and the AI benefit for a clinical decision-making task. 


\subsubsection{How and When HCPs appropriate AI?}

Addressing RQ2, participants’ accounts suggest that the perceived value of AI support depended partly on the fit between case demands and their existing expertise. For example, some participants turned to AI when encountering information outside their immediate expertise -- unfamiliar medications or pharmacology --, whereas others reported little need for AI when they already possessed the knowledge required for the case. Our observations suggest considering whether AI provides support beyond what an HCP can readily draw on from their existing expertise and perceived confidence levels. Overall, across participants' interactions, we observed three recurring roles for AI: verifying existing interpretations, offloading information-processing work, and extending knowledge.

\paragraph{AI as a Verifier of Professional Reasoning}
We saw several indications that HCPs used AI as a verifier and perceived it could be as one. For instance P16 reported that when they asked for the AI to rate their suggested plan, and when they got a positive feedback, it made them feel more confident with their plan. However, increasing confidence through AI confirmation \cite{ayorinde_health_2024} is not necessarily equivalent to improving the underlying decision. That gives us a natural connection to evaluative AI , cognitive forcing, and forward-reasoning support~\cite{zhang2024beyond,miller_explainable_2023,bucinca_trust_2021}. In high-stakes sensemaking, it may be more valuable when designed to expose counter-evidence, alternative interpretations, or information or hypotheses -- through the concept of evaluative AI ~\cite{miller_explainable_2023} -- that the professional has not considered in their decision-making process.

\paragraph{AI as Cognitive Offloader}

From our qualitative interviews and chat logs, we observed that HCPs were mainly relying in the chatbot for extracting trends from graphs; summarizing activity; calculating heart-rate targets; identifying frequencies/durations; avoiding manual day-by-day inspection of physical activity levels to identify patterns. In summary, we saw evidence of HCPs outsourcing cognitive tasks to AI, those which they could do themselves, but more time-efficiently. AI may be especially effective at reducing the temporal cost of obtaining information that the professional could otherwise derive themselves. However, our analysis shows that less cognitive work does not mean better decisions and better quality. Moreover, prior work shows that case complexity shapes AI's impact~\cite{abreu_ai_2026}, but can potentially increase workload and reduce efficiency. We forsee, that AI could help with preparatory information work which occurs before consultation \cite{gholamzadeh_applied_2021, pakianathan_dataenabledDialogue_2026}. 

\paragraph{AI as Knowledge Extender}

Next we saw that HCPs were using AI as an on-demand knowledge resource to augment their intelligence~\cite{tankelevitch_tools_2025,clark_extended_1998}. From our chat analysis we observed that they specifically asked questions about medications and medication related side effects on physical activity plans. This was also supported by our qualitative interviews where HCPs mentioned about lacking pharmacology expertise and that the chat functionality was useful for bridging their knowledge gap (e.g. P1 quote: [``AI support is useful] for something that I do not master''). 

These findings relates to complementarity in Human-AI collaboration as mentioned by Hemmer et al.~\cite{hemmer_complementarity_2025}. In their work, Hemmen et al., explain that there needs to be thoughtful analyses of the capabilities of humans and AI when designing for Human-AI collaboration~\cite{hemmer_complementarity_2025}. In particular, the authors argue that AI should not be built to mimic human experts but rather to target complementary capabilities, by training AI models excelling for task instances where human decision-making fails~\cite{hemmer_complementarity_2025}. In our case, this could be more towards assisting HCPs with verifying, offloading and extending their capabilities. 

\paragraph{From AI supported decision-making to decision-generation} 

In the main part of the study with the exercise plan generation for the 4 patients, we saw that AI initially supports HCPs with data sensemaking reasoning and before leading to crafting of the exercise plans. With the prototype concept of the AI based exercise plan generator (Figure~\ref{fig:exgen}), we introduced a shift in the decision-support paradigm. Our AI based exercise generator uses the patient data and clinical information to generate the initial plan and then allow the HCP to modify them manually. The HCPs appreciated how it could improve efficiency by saving time and also serve as a starting point from which the HCPs can make modifications using text prompts and accept the suggestions.

However, they had concerns about AI hallucination, need for transparency and explainability of why a particular exercise was suggested, and potential over-reliance and blind acceptance of suggestions and automation bias. They suggested that human-in-the-loop is essential and that professionals should always make the final judgement. 

\label{sec:trust}
\subsection{Trust and Verification Issues}
Despite our observation of high trust in AI output scores throughout our study, and feedback by participants that AI responses regarding medication side-effects making them feeling confident and reassured, we observe key tensions in trust. Across the three modalities -- verification, cognitive offloading and knowledge extension -- when professionals use AI to cover an expertise gap, they may simultaneously become less capable of checking whether the AI output is correct.

Firstly, the HCP may have sufficient knowledge to evaluate the AI output putting them in a safe state. However, for a HCP with lower knowledge, and higher reliance on AI particularly for knowledge extension, this could be an issue. Secondly, for offloading a similar case happens. When a HCP with moderate ability knows how the task should be done but chooses not to do it manually, it is fine, but someone with lower ability could be at a vulnerable position in assessing the AI output. Finally, for knowledge extension, the HCP might be asking about a topic which they don't know and might have low ability to verify the AI output, and also not know what they don't know, putting them in a risky position. On the one hand it raises the question of whether professionals can appropriately evaluate AI precisely when they need it because it exceeds their own capabilities? On the other hand, these examples emphasise that the more AI moves from offloading known work toward extending professional knowledge, the more important provenance, uncertainty, verification support, and appropriate knowledge grounding become.

\subsection{Design Implications}
\paragraph{Improving verification}
AI support should be designed to challenge and evaluate rather than simply agree. This could be through showing alternatives options, missing evidence, conflicting evidence, thereby improving \textit{option awareness}~\cite{miller_explainable_2023} -- the analysis and understanding of various options and their relative trade-offs. Furthermore, the AI support can enhance individuals thought process by acting ‘antagonistically’ thereby challenging users’ thinking ~\cite{tankelevitch_tools_2025}.  However at the same time, careful consideration should be made for verification of AI output as this could in turn result in \textit{verification burden} as defined by Gong et al. \cite{gong_cognitive_2026} where automation creates additional monitoring demands that may exceed the cognitive savings provided.

\paragraph{Improving offloading and trust}
Viewed through distributed cognition~\cite{hollan_distributed_2000}, we reflect on HCPs perform cognitive offloading and allocate information-processing work between themselves and AI while preserving professional judgment. HCP participants used AI support primarily to reduce the temporal and computational cost of information sensemaking. They asked the system to extract trends from graphs, summarize activity, calculate heart-rate targets, and identify exercise frequencies and durations—tasks they could perform themselves but at greater effort.

Furthermore, to improve credibility of AI outputs, while AI offloads routine information processing, it should also preserve the HCP’s ability to reconstruct and evaluate the underlying representation. Therefore, AI summaries and text outputs should allow the HCPs the verify the source of truth in the charts and abstractions via provenance-links as proposed by Pakianathan et al. \cite{pakianathan_exploring_2026}.

This design supports Human-AI collaboration in which AI reduces the temporal cost of explicit information processing while HCPs retain responsibility for contextual interpretation and decision-making. Overall, as quoted by Yang et al. ~\cite{yang_unremarkable_2019} AI support should be invisible and should fit existing work rather than become another task.

\paragraph{Improving knowledge extension}
Enabling HCPs to augment intellectual capabilities beyond their capabilities requires the need to have indicators in the AI system such as uncertainty indicator (e.g. expressed through percentages), explainability of how the output was generated and provenance of the information source and in the case of medical knowledge, grounding responses based on reliable sources such as peer reviewed literature or clinical guidelines~\cite{hurt2025use}.

\paragraph{Improving decision generation}
While our study showed hints of confidence increase amongst HCPs with the use of AI, we did not observe any significant improvement in quality of plans. To reduce risk of overconfidence while relying on AI, we suggest that exercise plan generation should allow for active professional judgement by the HCP while providing \textit{forward reasoning support}~\cite{zhang2024beyond,zhang_forward_2021}. Zhang et al.~\cite{zhang_forward_2021} describe forward reasoning as a process where the users, in our case HCPs, can form the decision themselves with augmentation of AI tools. Inspired by the conceptual design support paradigm proposed by~\cite{zhang2024beyond} we envision such a tool to encompass scaffolding questions such as firstly, asking HCP for an overall goal, followed by identifying constraints such as medications and side effects, exercise preferences of patient and their physical abilities and restrictions, preferred days of workout for the patients, FITT components and personalized heart rate zones. Subsequently, the HCP could edit the generated plan. In terms of input mechanism, voice input or rough sketch on paper could be converted into text prompt which is later fed into the AI exercise generator. Doing so could allow HCPs to think deeper about their exercise plan recommendations while improving speed of task.

\paragraph{Improving AI-driven clinical workflows}
Calota et al.~\cite{calota_sensemaking_2025} suggests the concept of sense making through making where designers understand the clinical context better through prototyping to design better clinical decisions about systems \cite{calota_sensemaking_2025}. In our case, by analysing sensemaking processes of the participants we could improve the design of the CDSS overtime. For instance, HCP interactions with the conversational system could be logged to better understand their sensemaking and information needs to generate pre-consultation summaries thereby saving interaction costs during the actual consultation. Contextualing our study findings around chat query patterns, a high level of queries around medication side effects signifies an opportunitiy for generating pre-consultation summaries or suggestions to allow the HCP to have an overview of a patient's medication-related side effects before deciding on exercise recommendations.

\paragraph{Adapt AI Support to Professional Expertise}
AI support should accommodate differences in professional expertise rather than provide the same level of assistance to all HCPs. Prior work shows that expertise shapes how clinicians engage with and rely on AI~\cite{kuper_psychological_2025,zhu_augmenting_2026}. For less experienced HCPs, systems could provide greater scaffolding through \textit{forward reasoning} ~\cite{zhang_forward_2021} rather than immediately presenting answers which may reduce opportunities for experiential learning and skill development~\cite{heudel_artificial_2026}. For more experienced HCPs, AI could give concise outputs while making additional explanations, evidence, and alternative interpretations available on-demand, thereby supporting professional judgement without unnecessarily increasing information load.

Furthermore, systems could allow HCPs to adjust the degree of AI assistance, for example by expanding or collapsing explanations, requesting additional evidence, or enabling more guided reasoning when needed. Such adaptable support could enable AI to scaffold clinicians when required while preserving professional autonomy and opportunities to exercise clinical judgment.

Overall, while integrating AI support to offer verify, offload, extend and generate capabilities, it is necessary to preserve opportunities for professional reasoning and skill development, particularly as increasing reliance on AI may reduce opportunities for experiential learning and contribute to longer-term deskilling~\cite{heudel_artificial_2026}.

\section{Limitations and Future Work}

Our study was conducted in a simulated planning setting in which HCPs worked with real patient-donated longitudinal data but did not interact directly with the patients. Consequently, participants had access only to the information represented in the study materials and could not clarify patients' preferences, routines, or circumstances through conversation. Future studies could evaluate AI-supported sensemaking in settings where HCPs can incorporate both longitudinal patient-generated health data and information elicited directly from patients. We recommended participants to spend approximately 10 minutes per patient. Although this supported a consistent study procedure, it may not reflect the time available for exercise prescription in routine cardiac rehabilitation practice. Our sample was relatively small and homogeneous. The 26 participants were exercise professionals who were predominantly young and early in their careers, with prior experience with generative AI/LLMs tool. These characteristics may have increased their familiarity with AI-decision support tools, limiting generalizability to older, less technologically experienced healthcare professionals who adoption, trust and interaction preferences may differ.
 
We also inferred task-specific knowledge gaps from participants’ reported uncertainty and patterns of AI use rather than directly measuring their knowledge in areas such as medication and exercise interactions. We did not manipulate professional expertise or AI correctness, nor did we directly assess participants’ ability to identify erroneous AI outputs. Inter-rater agreement for expert plan ratings was low and we followed-up with the raters to investigate this. Their feedback indicated some variation in how they assessed the three evaluation domains; for example, one assessed plan clarity based on whether the provided information was sufficient to independently complete an entire week of training. Additionally the experts suggested that the constrained 10+-minute, spreadsheet-based planning task may also have limited the detail available for assessment, while a more explicit rating rubric and a wider response scale (e.g., 7-point LIKERT) may have enabled more consistent and differentiated ratings. At the same time, disagreement may partly reflect broader variation in exercise-prescription practice: previous work with 47 physiotherapists in Belgium found substantial variation in exercise prescriptions for patients with cardiovascular disease \cite{marinus_are_2024}. Future work should therefore further investigate how plan quality and personalization can be evaluated reliably while accounting for legitimate variation in professional judgment.

Finally, the AI-supported condition combined several forms of support, including chart-level summaries, a holistic summary, and conversational interaction. Our study therefore cannot determine the independent contribution of each component to the observed outcomes or patterns of use. Future studies could use ablation or factorial designs to compare these forms of support independently and in combination. Interaction logging and eye tracking could further reveal how HCPs allocate attention across raw visualizations, AI-generated summaries, and conversational responses, and when they transition between these resources during sensemaking.

Taken together, our findings provide an empirical foundation for designing AI-supported decision tools that help HCPs make sense of longitudinal patient-generated health data while preserving professional judgment. By examining how HCPs selectively used AI to verify interpretations, offload information-processing work, extend their knowledge, and generate editable starting points for action, our study highlights opportunities for AI to support different parts of the sensemaking and planning process rather than replacing professional expertise. Future systems can build on this selective complementarity by adapting support to the capabilities of the professional, the demands of the case, and the type of reasoning required. Moving toward real-world cardiac rehabilitation settings offers an important opportunity to investigate how such systems can support shared decision-making, longitudinal care, and more personalized physical activity planning.
\section{Conclusion}
AI support in physical activity planning integrating longitudinal patient data showed that the value was selective depending on professional capabilities and the demands encountered during decision-making.  Our findings show that AI does not uniformly improve exercise planning quality, but can selectively complement professional capabilities by helping HCP verify their reasoning, offload data interpretation, and access knowledge and generate initial editable recommendations. We characterize this pattern as selective complementarity: AI contributes most when its capabilities provide additional value relative to what the professional is capable of. Future AI-CDSS should be designed around selective complimentary support, while preserving professional autonomy.




\section*{Acknowledgments}
We used generative AI tools solely for language editing and to improve
the clarity and readability of text written by the authors. All content
was reviewed and approved by the authors.
\bibliographystyle{ACM-Reference-Format}
\bibliography{bibliography}

\clearpage
\appendix

\begin{figure*}[t]
    \centering
    \includegraphics[width=0.8\textwidth]{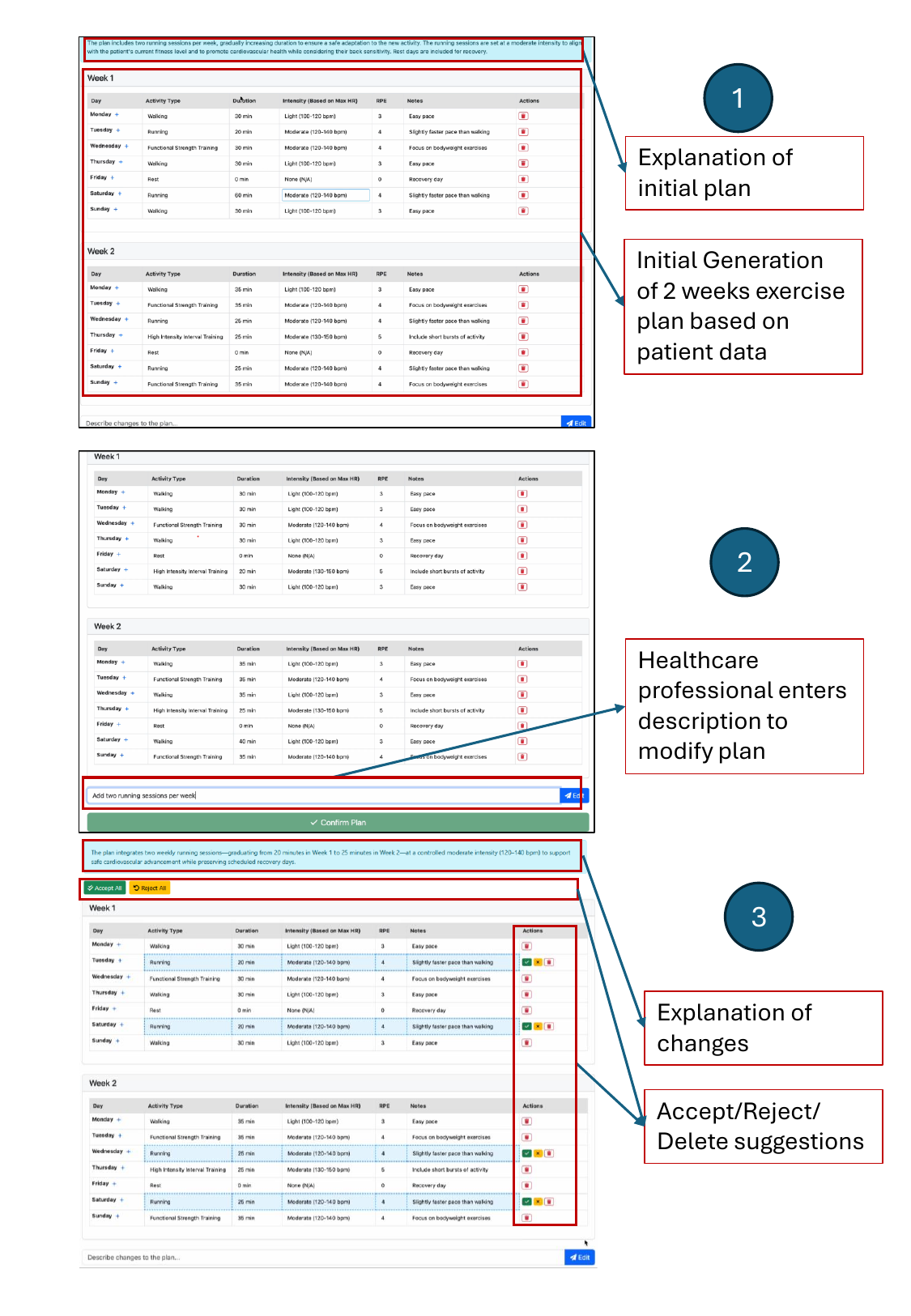}
    \caption{Workflow for AI-based Exercise Plan Generator}
    \Description{A three-panel sequential workflow outlining the interface for an artificial intelligence physical activity plan generator, labeled steps one, two, and three. Step one, labeled initial generation of two weeks exercise plan based on patient data with explanation, shows a two-week schedule presented as tables divided into Week one and Week two. Above the tables, a highlighted box provides a generated text rationale. Each row of the table details daily exercises with columns for Day, Activity Type, Duration, Intensity based on maximum heart rate, Rate of Perceived Exertion, Notes, and Actions. Step two, labeled add description to modify plan, shows the same two-week table schedule with a highlighted text prompt bar at the bottom containing the input text, Add two running sessions per week, next to an edit button and below a Confirm Plan button. Step three, labeled Accept, Reject, or Delete suggestions, illustrates the updated schedule where modifications appear as highlighted rows. Red bounding boxes highlight new options across the top bar to Accept All or Reject All, while individual row action columns display dedicated green checkmark, red cross, and trash icons next to modified entries.}
    \label{fig:exgen}
\end{figure*}

\section{Overview of AI-based Exercise Generator}
\end{document}